\documentclass[authoryear,review,12pt]{elsarticle}

\usepackage{amssymb}

\usepackage{soul}

\usepackage[colorinlistoftodos]{todonotes}
\usepackage{url}

\usepackage{natbib}
\journal{Beszédtudomány – Speech Science}

\usepackage{natbib}

\begin{document}

\begin{frontmatter}



\title{Towards Decoding Brain Activity During Passive Listening of Speech}


\author[BME]{Milán András Fodor}

\affiliation[BME]{organization={BME, TTK, DEPARTMENT OF COGNITIVE SCIENCE}}

\author[BMETMIT]{Tamás Gábor Csapó}
\author[BMETMIT]{Frigyes Viktor Arthur}

\affiliation[BMETMIT]{BME, VIK, DEPARTMENT OF TELECOMMUNICATIONS AND MEDIA INFORMATICS}

\begin{abstract}
The aim of the study is to investigate the complex mechanisms of speech perception and ultimately decode the electrical changes in the brain accruing while listening to speech. We attempt to decode heard speech from intracranial electroencephalographic (iEEG) data using deep learning methods. The goal is to aid the advancement of brain-computer interface (BCI) technology for speech synthesis, and, hopefully, to provide an additional perspective on the cognitive processes of speech perception.

This approach diverges from the conventional focus on speech production and instead chooses to investigate neural representations of perceived speech. This angle opened up a complex perspective, potentially allowing us to study more sophisticated neural patterns. Leveraging the power of deep learning models, the research aimed to establish a connection between these intricate neural activities and the corresponding speech sounds.

Despite the approach not having achieved a breakthrough yet, the research sheds light on the potential of decoding neural activity during speech perception. Our current efforts can serve as a foundation, and we are optimistic about the potential of expanding and improving upon this work to move closer towards more advanced BCIs, better understanding of processes underlying perceived speech and its relation to spoken speech.

\end{abstract}




\end{frontmatter}



\section{INTRODUCTION}

\subsection{Brain-Computer Interfaces and Deep Learning}

Brain-Computer Interfaces (BCIs) offer an exciting direction for direct communication between the human brain and external devices. Originally developed to assist individuals with neuro-motor disorders, BCIs have the potential to revolutionize a wide range of fields, including communication and rehabilitative technologies \citep{Brumberg_2011, luo2023brain}.

Recent advancements in deep learning have enabled considerable improvements in the interpretative power of BCIs. Deep learning, a subset of machine learning, involves artificial neural networks with multiple hidden layers, allowing for complex pattern recognition from high-dimensional data \citep{Schirrmeister_2017, Bashivan_2016}. As these deep learning techniques become more sophisticated, their application in BCIs is broadening, particularly in the field of communication BCIs, where one of the main goals is to reconstructing intelligible speech from neural activity~\citep{Akbari_2019}. However, significant challenges remain, particularly in less explored areas such as exploring the passive side of communication by decoding perceived speech, which is the primary focus of our research.

\subsection{The Cognitive Background of Listened and Spoken Speech}
The human speech process, both in speaking and listening, involves a multitude of complex cognitive processes. Neural signals generated during these processes hold rich information, which, if decoded successfully, could significantly enhance BCI technology for speech synthesis \citep{Hickok_2014, Pulvermuller_2014}.

Speech perception encompasses numerous processes such as acoustic analysis, phonetic and phonological processing, lexical access, and semantic comprehension \citep{Pei_2011, Herff_2015}. These processes are interconnected, often occurring in parallel, which leads to intricate neural representations of perceived speech within the brain \citep{Brandmeyer_Farquhar_McQueen_Desain_2013, Mesgarani_2014}.

Research into speech perception has revealed the involvement of several key brain regions. The superior temporal gyrus (STG) and the posterior superior temporal sulcus (pSTS) are particularly integral for processing acoustic features and phonetic components of speech \citep{Mesgarani_2014, Okada_2010}. These areas respond to various speech sounds and their characteristics, and their activation patterns often mirror the spectro-temporal dynamics of the incoming speech signal.

Beyond the acoustic-phonetic level, speech comprehension involves additional cognitive stages such as lexical access and semantic comprehension, which are associated with other brain regions. Wernicke's area, situated in the posterior part of the superior temporal gyrus, plays a significant role in understanding spoken language, linking the sound of speech to meaning \citep{Price_2012}.

\begin{figure}[!htbp]
\centering
\includegraphics[width=\textwidth, keepaspectratio]{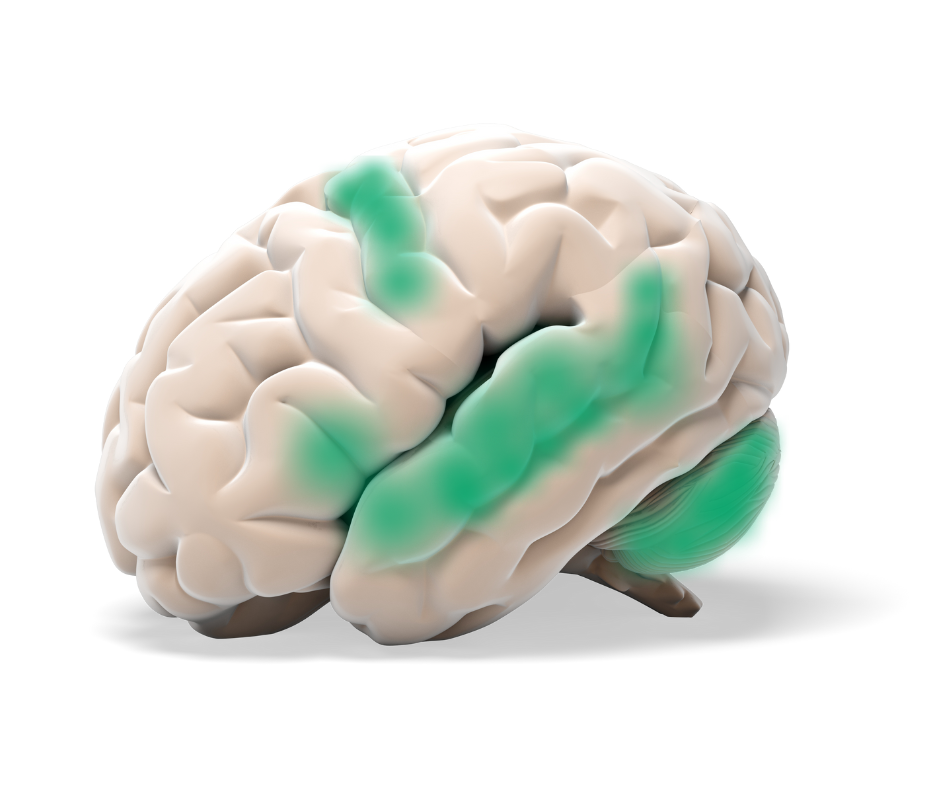}

\caption{Broca's area, the motor cortex, the cerebellum, Wernicke's area, and the superior temporal gyrus, posterior superior temporal sulcus highlighted as important areas of the brain regarding speech. Figure based on \citep{guenther2006cortical,hickok2007cortical, von2010human, Hein2008} .}
\label{agy}
\end{figure}

Interestingly, the perception of speech also engages brain regions traditionally associated with speech production. For instance, Broca's area, known for its role in speech production, also plays a part in speech perception, particularly when listeners are anticipating or predicting upcoming speech sounds \citep{Friederici_2011}. Similarly, activity in motor-related areas like the motor cortex and the cerebellum has also been observed during speech perception, potentially reflecting the listeners' internal simulation or mirroring of the speaker's articulatory movements \citep{eichert2020mapping, Lotte_2018}.

Key areas of the brain involved in speech perception are highlighted in Figure \ref{agy}.
These complex cognitive processes and their associated neural representations present both a challenge and an opportunity for BCI technology. Our research seeks to decode these intricate neural activities during speech perception to aid the advancement of BCI systems for speech synthesis, potentially enabling more naturalistic, communication-focused BCI technology.

\subsection{Speech Synthesis from neural activity}

Speech synthesis, the artificial production of human speech, is a rapidly evolving field that has undergone substantial advancements, particularly with the incorporation of deep learning and neural network methodologies \citep{Shen_2016, Oord_2016} next to regression-based approaches. These technological advancements have not only enhanced the intelligibility, naturalness, and expressivity of synthetic speech, but also allowed for the integration of complex neural data as an input source. 

Both neural network-based methods and traditional regression-based approaches, like those presented by \cite{pasley2012reconstructing}, have distinct advantages and disadvantages. Neural networks, particularly deep learning models, excel in handling complex, non-linear relationships in data, which can be crucial for accurately modeling the intricate patterns in auditory signals. They often achieve higher accuracy and can generalize better to new, unseen data. However, these models require large amounts of training data and substantial computational resources. On the other hand, traditional regression-based methods, while sometimes less accurate in complex scenarios, offer greater transparency and can be more interpretable. They are typically simpler to implement and require less computational power, making them more accessible for smaller-scale studies or applications with limited resources. Additionally, traditional methods can be more robust to overfitting when dealing with small datasets. Therefore, the choice between these approaches should be guided by the specific requirements and constraints of the speech reconstruction task at hand.

A key challenge lies in the adaptation of speech synthesis systems to real-world environments. Everyday communication often takes place amidst background noise, room reverberations, or with multiple speakers, conditions that can considerably impair the performance of conventional speech synthesis systems \citep{Godoy_2018}. Developing algorithms capable of effectively synthesizing speech under such challenging conditions is a critical area of ongoing research.

As the field of speech synthesis evolves, there is an emerging interest in faster, more accurate and more naturalistic approaches. One possible avenue to get closer to this goal is could leveraging BCI to decode heard speech from neural activity \citep{Pei_2011, Brandmeyer_Farquhar_McQueen_Desain_2013}. This innovative approach would allow for the synthesis of speech that the user hears, rather the user's input. When further developed, this method could potentially assist in instantly storing information we consciously perceive. Additionally, it may enhance overt speech synthesis, as we hear our own speech.

\section{RESEARCH OBJECTIVE}

This study seeks to employ a dataset of iEEG recordings collected during passive listening of speech. Utilizing deep learning algorithms, we aim to construct a model that aspires to decode the heard speech from these neural activities. By doing so, we anticipate contributing to advancements in BCI technology and enhancing our theoretical understanding of cognitive speech processing.
The adoption of these advanced computational techniques could enable us to unravel the intricate neural representations of perceived speech, and these insights pave the way for advancements in BCI systems \citep{Schirrmeister_2017, Bashivan_2016}.
The adoption of iEEG data is particularly advantageous due to its high signal-to-noise ratio, enhanced spatial resolution, and ability to capture a broad range of frequency bands, making it highly suitable for speech decoding \citep{Halgren_2019, Crone_2001}.

We articulate our decision to situate our research in the context of passively listened speech. While existing BCI research often emphasizes speech production, the area of listened speech remains relatively unexplored. We propose that this angle harbors untapped potential, promising novel insights into the cognitive dimensions of speech processing and a fresh angle for speech decoding efforts \citep{Pei_2011, Brandmeyer_Farquhar_McQueen_Desain_2013}. There may be certain disabilities where brain damage affects auditory processing in such a way that, although neural representations of heard sounds are present, they are not fully perceived by the individual. While recording sound is an obvious solution, this technology, when fully developed, could offer an instantaneous alternative that might only record sounds the individual focuses on. It also has implications for overt speech decoding, since we hear our own words, when speaking.

In conclusion, this study represents an effort to elucidate the complex relationship between speech perception and production, and their neural representations, while advancing the development of naturalistic, communication-focused BCI technology.

.

\section{METHODS}

\subsection{Dataset}

This research uses the 'Open multimodal iEEG-fMRI dataset' \citep{Berezutskaya2022}, a publicly available resource that combines iEEG with fMRI data. The high spatial and temporal resolution of the dataset offers detailed insights into speech and language processing.

\subsubsection{Participants}
The dataset contains data from fifty-one Dutch epilepsy patients undergoing diagnostic procedures at the University Medical Center Utrecht. Of these, sixteen provided written consent for their clinical data to be used for research, out of which we choose the most suited ones. The patients' ages varied, with an average of 25 and a standard deviation of 15, including 32 females. For patients under 18, consent was obtained from their parents or legal guardian. The study was approved by the Medical Ethical Committee of the University Medical Center Utrecht, in line with the Declaration of Helsinki (2013).

\subsubsection{Experimental Procedures}
The patients participated in two main types of experiments: movie-watching and resting state. The movie-watching experiment, which involved the patient watching a short film, was part of the standard battery of clinical tasks for presurgical functional language mapping. The resting state experiment, which required the patients to rest for three minutes, was conducted for research purposes. For those patients who did not participate in a separate resting state task, a 3-minute 'natural rest' period was selected from their 24/7 clinical iEEG recordings.

\subsubsection{Stimuli}
The stimulus for the movie-watching experiment was a 6.5-minute short movie composed of fragments from "Pippi on the Run" (P\aa{}rymmen med Pippi L\aa{}ngstrump, 1970). The movie was edited to form a coherent plot and consisted of 13 interleaved blocks of speech and music, each 30 seconds long. The movie was originally in Swedish but dubbed into Dutch. Detailed annotations of the audio and video content of the movie stimulus were provided.

\subsubsection{Electrode Implantation}
Electrode types varied based on clinical requirements. Forty-six patients had ECoG grids with 48 to 128 contact points. Six patients had high-density ECoG grids with 32 to 128 contact points. Sixteen patients had sEEG electrodes with 4 to 173 contact points. Most electrodes covered perisylvian areas and frontal and motor cortices.

\subsubsection{Data Acquisition}
Intracranial EEG (iEEG) data were acquired using a 128-channel recording system (Micromed, Treviso, Italy) during the experimental tasks. The majority of patients' data were sampled at 512 Hz and filtered at 0.15--134.4~Hz, while in some cases, the data were sampled at 2048 Hz and filtered at 0.3--500~Hz. An external reference electrode was used for signal referencing, typically placed on the mastoid part of the temporal bone. Besides, six patients had their HD ECoG data recorded either simultaneously with the clinical channels or in separate sessions.

\subsection{Data Availability}

The dataset can be accessed at: \url{https://openneuro.org/datasets/ds003688}. To maintain confidentiality, identifiable information and individual MRI scans have been removed. The order of subjects in the dataset has been randomized to further ensure anonymity.

\subsection{Data Validation}

Preprocessing of the iEEG data was carried out using MNE-Python (\url{https://mne.tools}). 

To ensure data quality, the subjects' neural activity during speech and music blocks was compared  by the team behind the dataset. \citep{Berezutskaya2022}.

\subsection{Prime Subjects}

To facilitate the most effective and meaningful analysis for this study, we utilized a rigorous selection process for the subjects, oriented primarily around a key determinant: the level of correlation that each subject demonstrated with the speech envelope during the movie, as noted by the team who compiled the dataset \citep{Berezutskaya2022}. 

\begin{figure}[!h]

\centering

\includegraphics[width=\textwidth, keepaspectratio]{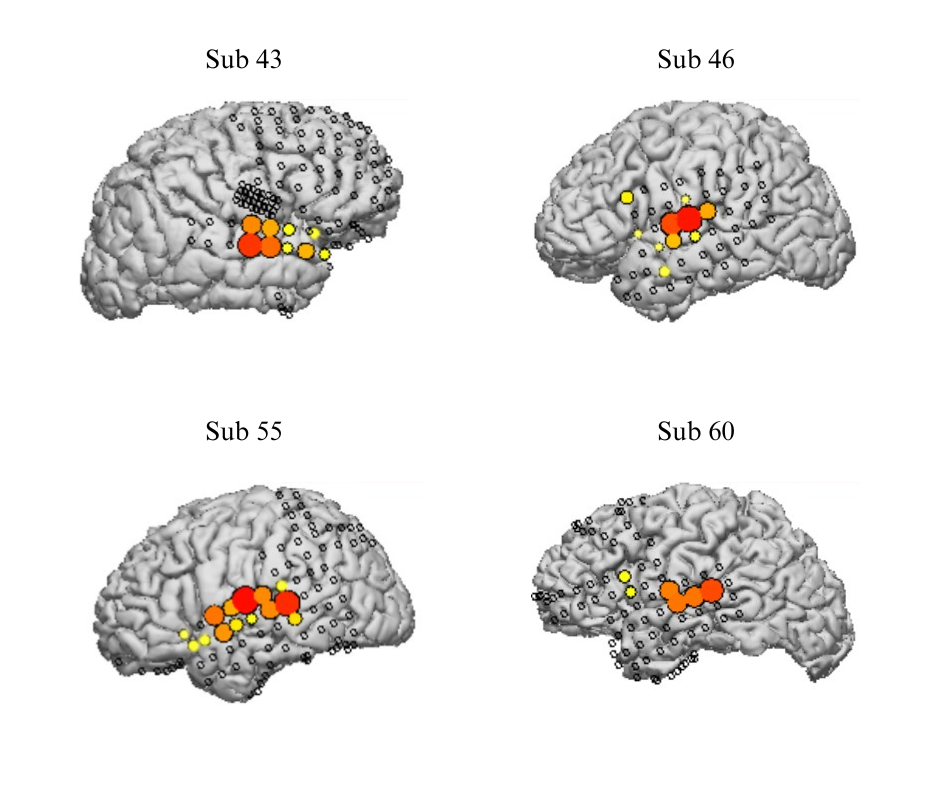}

\caption{The four subjects with the highest correlation with the speech envelope. From \citep{Berezutskaya2022}.}

\label{bestbrains}

\end{figure}

Additionally, our selection strategy was influenced by the need to optimize our limited time and GPU resources. This subject selection methodology stemmed from the hypothesis that individuals whose neural activity closely mirrored the dynamic ebb and flow of the speech envelope would be ideal candidates for this study. From the pool of potential subjects, four individuals were eventually selected, as shown in Fig. \ref{bestbrains}. 

These participants displayed notably high correlation values, likely stemming from the placement of intracranial electrodes covering key areas associated with speech perception and production, including the Broca’s area, the motor cortex, the cerebellum, Wernicke’s area, and the superior temporal gyrus. The selection process ensured the recruitment of subjects whose neural responses would yield the richest and most insightful data for decoding and reconstructing speech from neural signals.

\begin{figure}[!htbp]

\centering

\includegraphics[width=\textwidth]{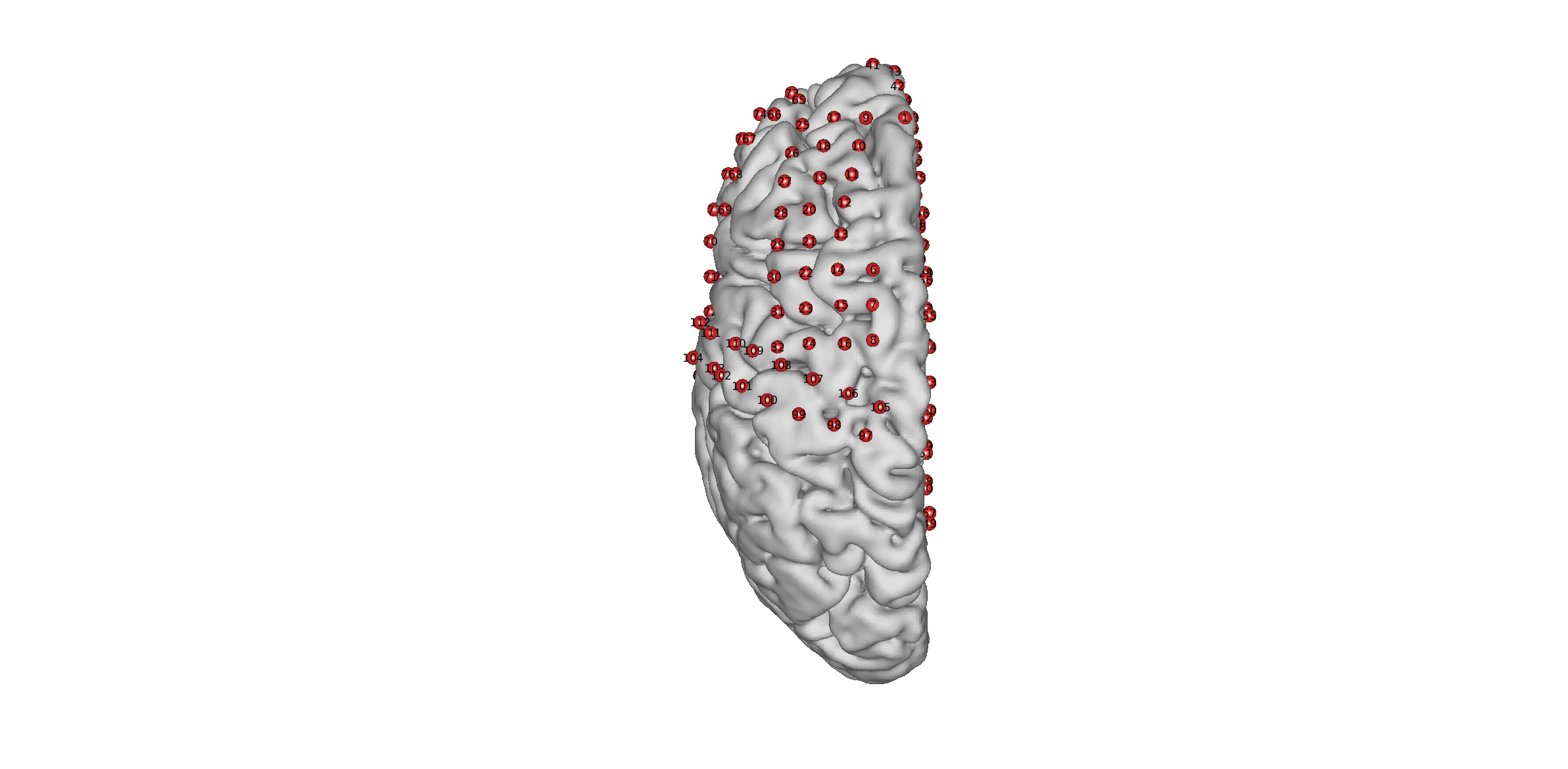}

\caption{The electrode positions for Subject 38. Extracted from the Open multimodal iEEG-fMRI dataset \citep{Berezutskaya2022}}

\label{38electrodes}

\end{figure}

In addition to data-driven selection, manual selection ensured coverage of essential brain regions. In particular, Subject 38 was chosen for their exceptional coverage of electrodes over the motor cortex, the Broca’s area, and the superior temporal gyrus — see Fig.~\ref{38electrodes}. This unique electrode placement may offer a unique opportunity for more accurate and nuanced speech reconstructions.

By employing both quantitative and qualitative selection criteria, we identified the subjects who were most likely to contribute valuable data to the study.


\subsection {Preparing Data for training}

\begin{figure}[!h]

\centering

\includegraphics[width=\textwidth, keepaspectratio]{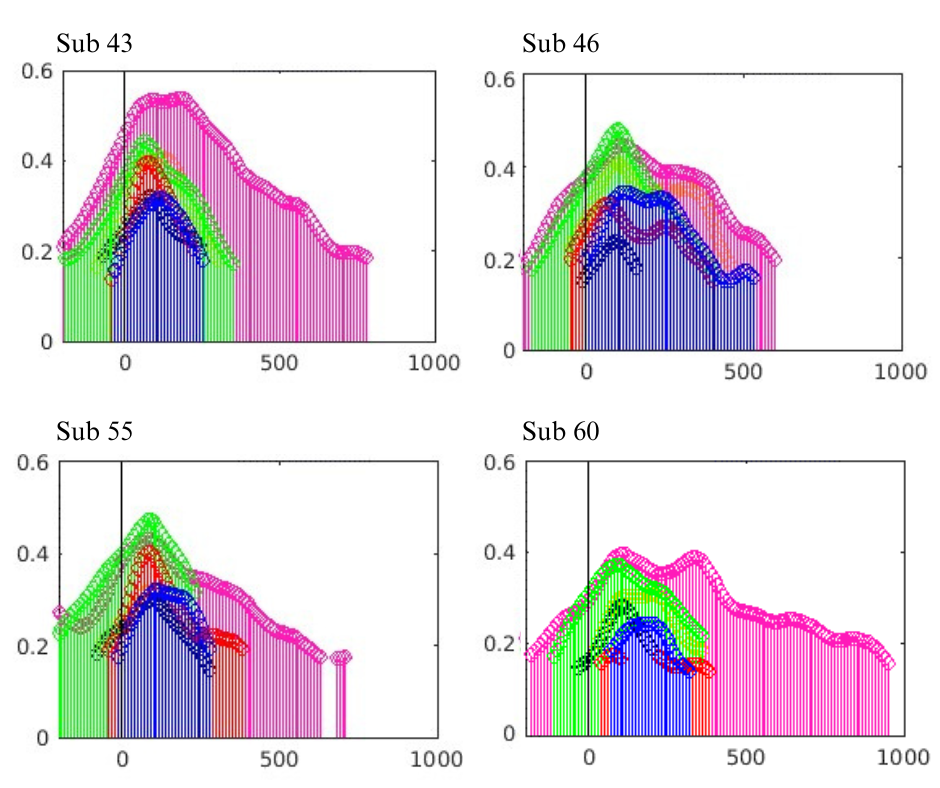}

\caption{Lagplots of the cross-correlation of the electrode’s high-frequency band signal and the sound envelope \citep{Berezutskaya2022}.}

\label{shift}

\end{figure}

\subsubsection{Audio and Lag Correction}

The audio data is first loaded using the librosa library. Figure \ref{shift} provides a visual representation of the cross-correlation between the electrode’s high-frequency band signal and the sound envelope. Different colors correspond to different 30-second speech blocks. The observed average delay is approximately 150 milliseconds, which we accounted for by shifting the audio backwards by 150 milliseconds, thereby enhancing the alignment of the decoded speech with the original auditory stimulus.

For mel-spectrogram estimation from speech, 80 bins were used using librosa mel-filter defaults.  Essential STFT parameters were set, including a filter length of 1024, a hop length of 10 ms, and a mel frequency range spanning from 0 to 8000 Hz, 80 frequency bins. The sampling rate was 22050 Hz.

\subsubsection{Filtering and cropping only speech segments of iEEG}

All subjects' brain signal data were sampled at 512 Hz. Initially, ,,ECoG'' and ,,sEEG'' type channels were selected, and defective channels were removed. A notch filter was applied to counter line noise at 50 Hz and its harmonics. The data was then re-referenced using the common average signal. 

We extracted EEG data corresponding to the segments where speech was present in the movie. This was achieved by selectively slicing the raw\textunderscore{}car data (the preprocessed EEG data) and the mel\textunderscore{}data (mel-spectrogram estimated from the speech stimuli) based on provided annotations, thereby focusing the analysis on the brain’s response to auditory speech stimuli. The result of this procedure was a refined set of data (raw\textunderscore{}car\textunderscore{}cut and mel\textunderscore{}data\textunderscore{}cut), encapsulating the EEG responses to speech stimuli, thus enhancing the relevance and accuracy of the subsequent deep learning model training.

\subsubsection{Feature extraction}

In the feature extraction process, we first apply linear detrending to the EEG data, effectively removing linear trends and reducing potential artifacts. The data is then segmented into overlapping windows, each defined by a specific length (0.05 ms) and shift (0.01 ms). Within these windows, we perform bandpass filtering, specifically targeting the 1--120 Hz frequency range, as everything from a delta to high gamma frequencies are relevant to speech, when we are also interested in speech perception \citep{lopez2022state}. Subsequently, the Hilbert transform is applied to the filtered data to derive the analytic signal, enabling us to calculate the amplitude envelope. The final step involves computing the mean amplitude of this envelope for each window across all EEG channels. The resulting output is a 2D feature matrix, where each row represents a time window and each column corresponds to an EEG channel. This matrix encapsulates the mean amplitude of the target frequency band for each window and channel, providing a concise representation of the EEG data for further analysis.

\subsection{Deep Learning Training}

Deep learning, renowned for its efficacy in abstract pattern extraction from extensive high-dimensional data sets, is a natural fit for parsing intracranial electroencephalogram (iEEG) data. Notably, its prowess in related tasks motivated its selection.

\begin{figure}

\centering

\includegraphics[width=\textwidth]{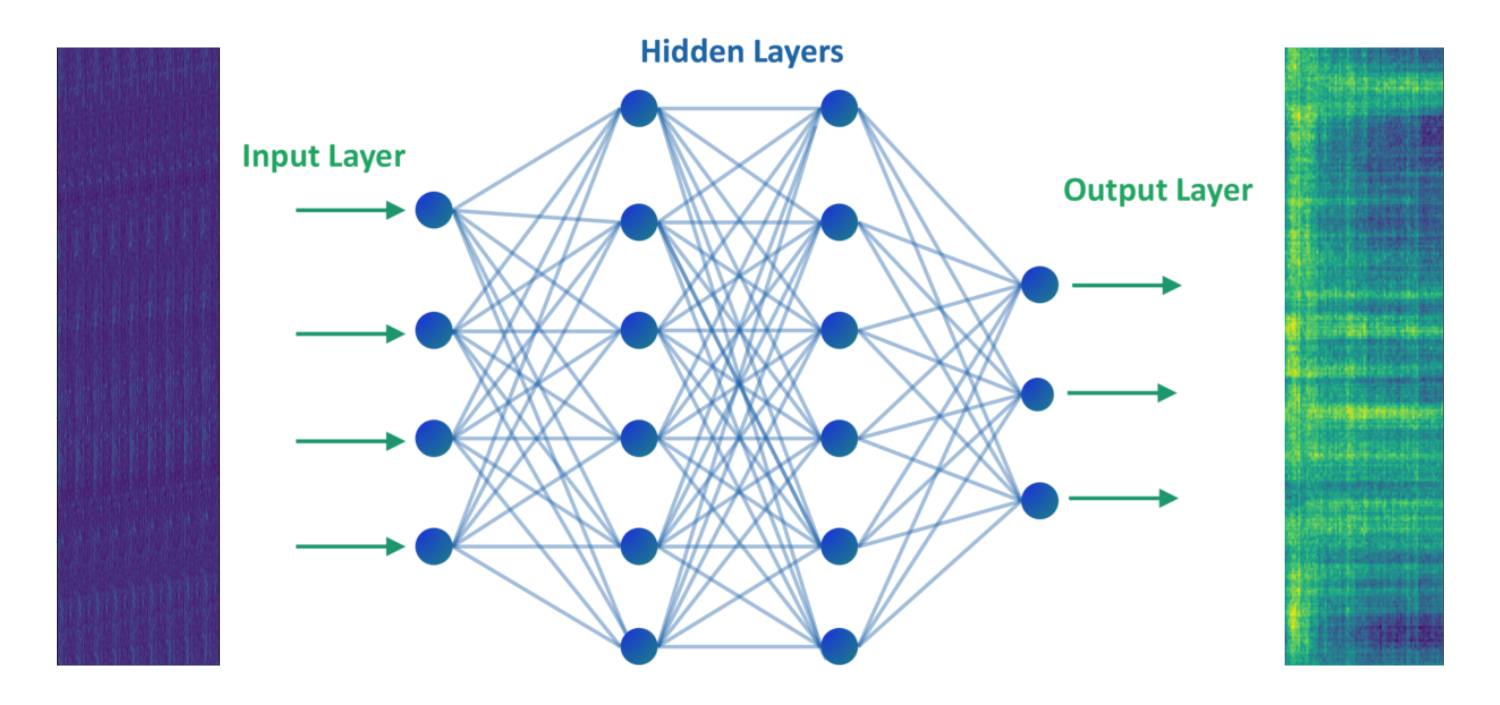}

\caption{Visual representation of the iEEG input and mel-spectrogram output of the DNN. }

\label{archi}

\end{figure}

We utilized Fully Connected Deep Neural Networks (Fc-DNNs) and 2D Convolutional Neural Networks (2D-CNNs) for this research, chosen after assessing their inherent properties and suitability for predicting mel-spectrograms from iEEG data. Figure \ref{archi} presents a simplified illustration of the transformation process: iEEG inputs being fed into the DNN architecture, and subsequently producing mel-spectrogram outputs. The selection process was iterative, involving comprehensive evaluation of multiple model architectures, training strategies, and optimization techniques. The configurations delivering optimal performance were chosen for the final models.

To ensure transparency and repeatability, all code, files, and scripts utilized for data preprocessing, model training, and result analysis are publicly shared at:\ (WARNING: clicking this link might reveal author identities)

\url{https://github.com/MILANIUSZ/speech2brain2speech}.

The training process employed an RTX 3070 GPU and an AMD Ryzen 5 3600 processor, with the environment set up using Docker and the public image “thegeeksdiary/tensorflow-jupyter-gpu”. This setup ensured efficient hardware utilization for model training and evaluation.

\subsubsection{Fc-DNN}

Fully Connected Deep Neural Networks, also known as Multilayer Perceptrons (MLPs), are versatile neural networks used extensively for regression tasks. This study employed an Fc-DNN model with one hidden layer of 3000 neurons. This configuration was systematically chosen after multiple iterations to ensure optimal performance while avoiding overfitting, as increasing model complexity didn’t substantially improve accuracy but led to overfitting.

The Rectified Linear Unit (ReLU) was used as the activation function for the input layer due to its ability to handle the vanishing gradient problem, and a linear activation function for the output layer, fitting for a regression task. Adam optimizer was used due to its efficiency.

Data was partitioned into training, validation, and test sets (80\%, 10\%, 10\%, respectively).  EEG and mel-spectrogram data were scaled using MinMaxScaler and StandardScaler, respectively. The Fc-DNN model was built using Keras with a hidden layer of 3000 neurons (ReLU activation) and an output layer of 80 neurons (linear activation). The model was compiled with Mean Squared Error as the loss function, trained for a maximum of 50 epochs with a batch size of 32, with early stopping for overfitting prevention.



\subsubsection{2D-CNN}

Two-Dimensional Convolutional Neural Networks excel at grid-like data processing tasks. For this study, a 2D-CNN was used to process the spectrogram data obtained from EEG recordings, with an 80\% allocation for the training set (similarly to Fc-DNN). Both the input and output data were normalized using the mean and standard deviation from the training set. 

The 2D-CNN model architecture consisted of three convolutional layers with ‘swish’ activation function and dropout layers to prevent overfitting. Padding was applied to the input so that the output has the same length as the original input when the stride is 1. The model included a max pooling layer for dimensionality reduction, followed by a flatten layer and a dense layer with ‘swish’ activation. The output layer was a dense layer with a linear activation function, aligned with the training spectrogram shape. The model was compiled with the ‘Adam’ optimizer and the ‘mean squared error’ loss function, with training conducted over 100 epochs with a batch size of 128. Overfitting prevention was handled through early stopping and learning rate adjustment.

After training, the predicted spectrogram was inverse transformed to its original scale and saved for subsequent evaluation.

For detailed training parameters of the neural network, please refer to the supplementary materials and our GitHub repository.(WARNING: clicking this link might reveal author identities) \url{https://github.com/MILANIUSZ/speech2brain2speech}. 



\subsubsection{Evaluation Methods}

The performance of the Fc-DNN and the 2D-CNN was evaluated using Mean Squared Error (MSE) as the measure, with lower MSE indicating better mel-spectrogram prediction from EEG signals. Training was conducted 10 times for each model.

For qualitative assessment, the predicted, scaled mel-spectrograms were plotted in comparison to the orginial test mel-spectrograms. The discrepancies offered insights into the models’ performance.

Subject 13 was selected as a baseline because the implanted electrodes primarily covered areas on the occipital lobe. The occipital lobe is hypothesized to have fewer associations with the cognitive processes involved in perceived speech. Therefore, this choice provides a meaningful reference point for our analysis.

Additionally, an informal auditory evaluation was done by the first author. The reconstructed mel-spectrograms were converted back into audio signals using using the Griffin-Lim algorithm, implemented through the librosa library in Python, allowing aural comparisons of original and synthesized signals, revealing potential model shortcomings.

\section{RESULTS}

\subsection{Fully-connected Deep Neural Network}

The Fc-DNN was trained on the data from 6 subjects ( four(s43, s46, s55, s60) prime subjects, one "ideal electrode placement" subject(s38) and one not ideal electrode palcement subject(s13)), and the performance of the model for each subject is summarized in Table \ref{tab:Fc-DNN_Results}. The table presents the best training loss and validation mean squared error (MSE) achieved for each subject.


\begin{table}[h]

\centering

\begin{tabular}{|c|c|c|}

\hline

\textbf{Subject} & \textbf{Best Training Loss } & \textbf{Best Validation MSE} \\

\hline

38 & 0.0210 & 0.6982 \\

43 & 0.0336 & 0.7381 \\

46 & 0.2643 & 0.7923 \\

55 & 0.2015 & 0.7210 \\

60 & 0.3900 & 0.6520 \\

13 & 0.3256 & 0.8052  \\ 

\hline

\end{tabular}

\caption{Performance of the Fc-DNN for each subject.}

\label{tab:Fc-DNN_Results}

\end{table}

The training loss values represent how well the model is able to predict the mel-spectrogram data from the EEG signals during training. Lower training loss indicates a better fit of the model to the training data. The validation MSE, on the other hand, provides a measure of the model’s performance on unseen data, with lower MSE values representing better generalization performance.

From Table \ref{tab:Fc-DNN_Results}, it can be observed that the model achieved the lowest training loss with subject 43, indicating the model was able to fit the training data most effectively for this subject. On the other hand, the model demonstrated the best generalization performance on unseen data with subject 60, as indicated by the lowest validation MSE.

\subsection{Two-Dimensional Convolutional Neural Network}

Just like the Fc-DNN, the 2D-CNN model was trained on the data from the six different subjects. The performance metrics for the 2D-CNN, specifically the best training loss and the validation mean squared error (MSE) for each subject, are outlined in Table \ref{tab:2D-CNN_Results}.

\begin{table}[h]

\centering

\begin{tabular}{|c|c|c|}

\hline

\textbf{Subject} & \textbf{Best Training Loss} & \textbf{Best Validation MSE} \\

\hline

38 & 0.4121 & 0.7023 \\

43 & 0.5321 & 0.7326 \\

46 & 0.8043 & 0.6920 \\

55 & 0.9605 & 0.7879 \\

60 & 0.9039 & 0.7922 \\

13 & 0.9039 & 0.8781 \\ 
\hline

\end{tabular}

\caption{Performance of the 2D-CNN for each subject.}

\label{tab:2D-CNN_Results}

\end{table}

The 2D-CNN model performance is evaluated using the same metrics as the Fc-DNN model: the training loss, the validation MSE and informal listening to the synthesized audio. In this case, subject 46 achieved the lowest validation MSE.

\subsection{Mel-spectrogram demonstration samples}

\begin{figure}

\centering

\includegraphics[width=\textwidth, keepaspectratio]{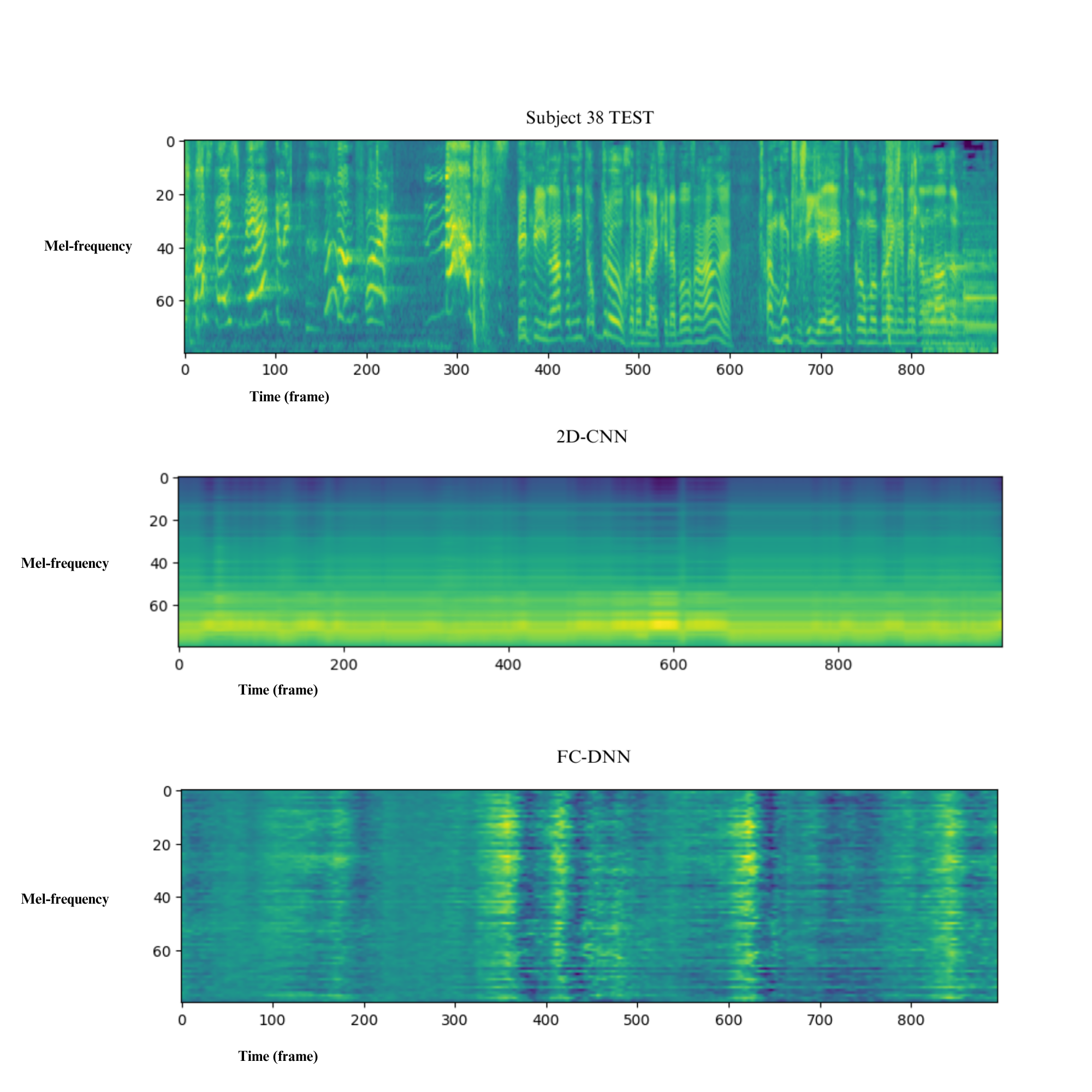}

\caption{Mel-spectograms for subject 38.}

\label{melspec38}

\end{figure}

Fig.~\ref{melspec38} shows an original speech stimuli sample (top) and those mel-spectrograms generated from iEEG input by the 2D-CNN (middle) and Fc-DNN (bottom) networks. Based on visual inspection, we can see that the result of 2D-CNN is oversmoothed, whereas the FC-DNN was able to generate more “realistic” patterns. However, the similarity between the original audio stimuli and the predicted spectrogram is still not satisfactory. 

\begin{figure}

\centering

\includegraphics[width=\textwidth, keepaspectratio]{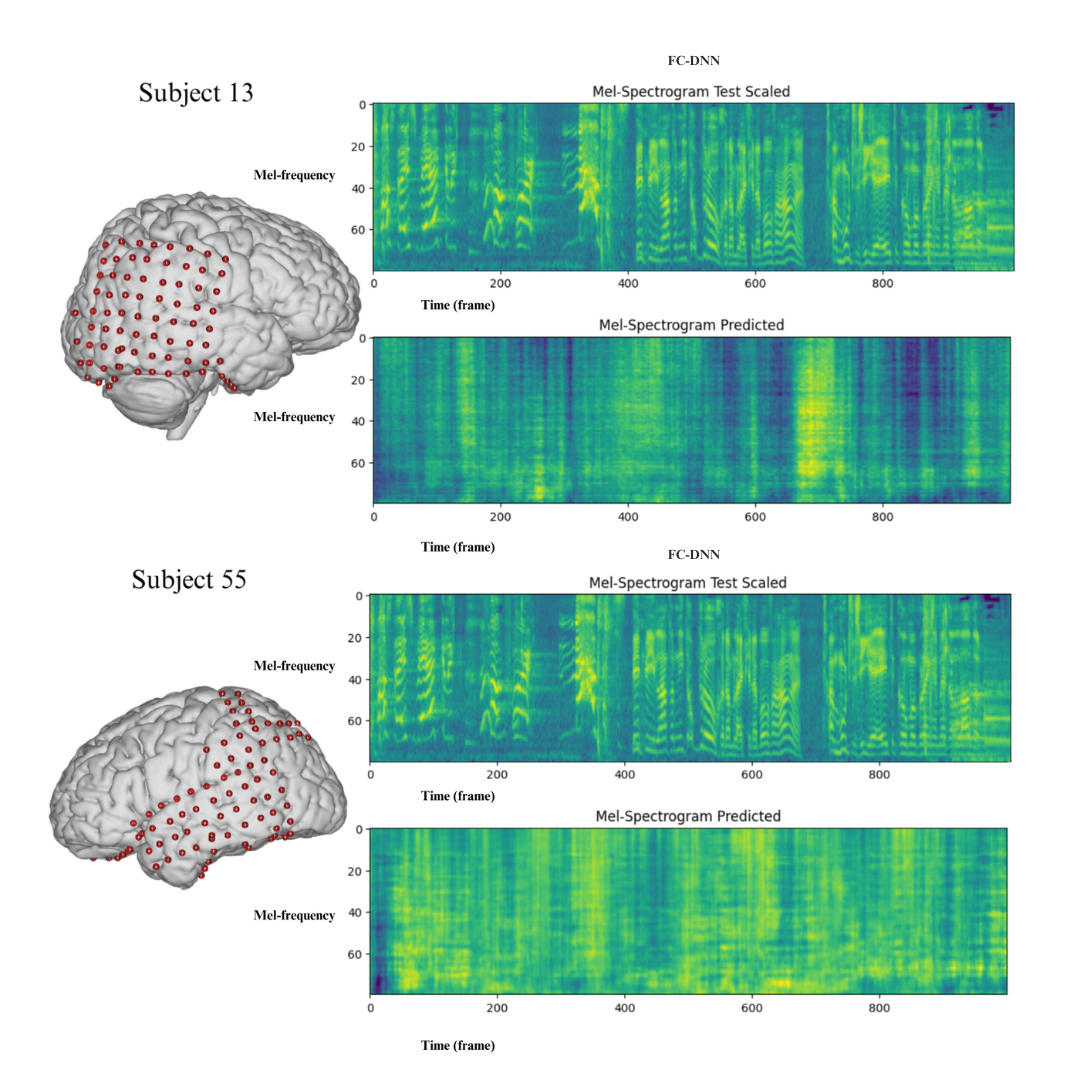}

\caption{Electrode placement and mel-spectrograms (based on the FC-DNN) comparison for subjects 13 and 55.}

\label{melspec13_55}

\end{figure}

Fig.~\ref{melspec13_55} compares the iEEG-to-speech results of two subjects. Based on this, the results of subject 55 seem to be more realistic, probably because his electrodes are located at more relevant areas of the brain. 

\subsection{Audio synthesis}

Both models’ synthesized audio underwent informal human evaluation by the first author, to assess its quality and intelligibility. While the speech wasn’t comprehensible, in some cases, the model captured some auditory elements, such as the silences. However, the accurate reconstruction of speech content remains a huge challenge.

\section{DISCUSSION AND WAY FORWARD}

\subsection{Speech Decoding}

The selected deep learning architectures, Fc-DNN, and the 2D-CNN, especially in light of the limited training data, have demonstrated the potential of the approach by finding patterns in perceived speech’s neural activity indicated by the reduction of test loss in a consistent manner. 

Despite the demonstrated potential of this approach, significant challenges remain with the methodology. A noteworthy issue in the current study is the difficulty in achieving satisfactory accuracy levels for both validation and test sets concurrently, despite multiple training iterations. The resulting mel-spectrograms, although indicating some pattern recognition and learning within the data, failed to provide a realistic spectrogram. Consequently, the audibility and clarity of the synthesized speech generated from these mel-spectrograms were low.

Previous studies, such as \cite{anumanchipalli2019}, which synthesized intelligible speech from neural activity recorded during active reading tasks, reported successful outcomes. However, our study primarily relies on passive tasks, which don’t generate robust motor and auditory brain responses like active tasks. Therefore, the differences between the results obtained in their study and ours can be attributed to the contrasting nature of the tasks involved.

At the same time, our study resonates with other research, such as \cite{akbari2019}, which aimed to decode spectrograms from brain activity recorded during passive listening tasks. Their findings, which reported challenges in generating realistic spectrograms and clear synthesized speech, echo the issues encountered in our study.

However, we must exercise caution when comparing these studies due to methodological variations, such as data collection techniques, preprocessing steps, model architectures, and evaluation metrics. For instance, some studies might employ invasive electrocorticography (ECoG) for data collection, resulting in high-resolution data, while others might utilize non-invasive methods like EEG or fMRI.

Despite these variations, the overall trend underscores the complexity of speech decoding, especially during passive listening scenarios, and highlights the need for more careful data preparations and/or significant technological advancements for reliable synthesis of clear speech from such brain activity.

\subsection{Cognitive Conclusions}

Upon comparing the accuracy and spectrograms, it seems that the patients with electrode placements, which were hypothesized to yield better results based on the literature, shown in Fig.~\ref{bestbrains}, do indeed show improved outcomes. As illustrated in Fig.~\ref{melspec13_55}, a disparity in performance can be observed between subject 13 (for which we achieved validation MSE of 0.805 with the FC-DNN and 0.878 with the CNN), serving as our baseline, and subject 55. The latter's electrode placement is more closely aligned with regions typically associated with speech processing, thereby reinforcing the crucial role of electrode placement in the accurate prediction of perceived speech.


These findings also hint at the possibility of shared characteristics in neural activity during passive listening and spoken speech, which might align with theories such as the ‘motor theory of speech perception’ \citep{Liberman1985, galantucci2006motor}, the ‘neural reuse’ theory \citep{Anderson2010} or the role of ‘mirror neurons’ in speech \citep{Rizzolatti2008}. However, these connections should be interpreted with caution, as our study does not provide definitive evidence for such theories.

\subsection{Limitations and Future Directions}

The big limiting factor of our study's success was the alignment of iEEG and audio data. It is challenging, and also amplified by the limitation of the dataset size. Future endeavors should focus on improved synchronization methods, larger, more diverse datasets, and the utilization of more advanced neural network architectures, e.g. transformer-based methods which can better handle temporal misalignment. In addition, including audible speech reproduction scenarios and interpretability techniques for neural networks could offer deeper insights into cognitive processes. While our study focused on intracranial EEG data, future research may consider other modalities like MEG or fMRI for more comprehensive data. 

Moreover, an interesting avenue for future work could be the integration of multi-modal data, such as neural activity from various brain regions, and additional data sources like facial movements, articulatory gestures or visual cues \citep{DBLP:journals/corr/abs-1904-05259, DBLP:journals/corr/abs-2104-14467,Csapo2023c}. This approach could help enhance the decoding performance and accuracy of speech BCIs.

\subsection{Future BCI}

The advancement of communication BCI continues, we try to create systems that work more accurately, faster and in a more naturalistic way. However, despite all the advancements in the field, challenges remain. Current neural recording techniques, such as invasive iEEG, offer high resolution but are impractical for widespread use. There is also a demand for even more efficient, speech-specific decoding algorithms, as existing models can require extensive datasets and substantial computational resources. Further, the field might benefit from a deeper understanding of speech processes in the brain.


This study tried to emphasize the potential role of perceived speech in the field. Our current efforts can serve as a foundation, and we are optimistic about the potential to expand and improve upon this work, moving closer to more advanced and effective BCIs.

\section{Acknowledgements}

We would like to thank the authors of the 'Open Multimodal IEEG-FMRI Dataset' for making the data available.

Projects and grants to be added after reviews.

\appendix


\bibliography{cas-refs}





\end{document}